\documentstyle[prl,aps]{revtex}
\begin{document}
\draft
\title{\hfill{\rm {CERN-TH/2002-072}}\\
\hfill{\rm {hep-ph/0204217}}\\
\hfill{\rm {April, 2002}}\\
{\bf {Symmetries, Large Leptonic Mixing and a Fourth Generation}}}
\author{Joaquim I. Silva-Marcos\cite{juca}}
\address{Theoretical Physics Division, CERN, \\
CH-1211 Geneva 23, Switzerland}
\maketitle
 
\begin{abstract}  
We show that large leptonic mixing occurs most naturally in the framework of
the Sandard Model just by adding a fourth generation. One can then
construct a small $Z_4$ discrete symmetry, instead of the 
large $S_{4L}\times S_{4R}$,
which requires that the neutrino as well as the charged
lepton mass matrices be proportional to a $4\times 4$ democratic mass
matrix, where all entries are equal to unity. Without considering the
see-saw mechanism, or other more elaborate extensions of the SM, and
contrary to the case with only 3 generations, large leptonic mixing is
obtained when the symmetry is broken.
\end{abstract}   

\section{Introduction}

Recently \cite{novi}, it was suggested that the existence a fourth
generation of fermions is in agreement with the latest electroweak precision
data. However, this simple extension of the Standard Model (SM) could also
be a natural way to explain the smallness of the masses of the first 3
generations of neutrinos. It was shown \cite{juca4neut}, for the lepton
sector and within a democratic weak basis, that small masses for the first 3
neutrinos are compatible with large mixing angles, assuming that neutrinos
are of the Dirac type. In the democratic $4\times 4$ limit, all Yukawa couplings 
between the different generations of leptons are equal, and 
only the fourth generation acquires mass, 
for each lepton sector. 
Democracy provides an alternative explanation 
(e.g. to the see-saw mechanism) for the smallness of the neutrino masses as it 
does not require very small Yukawa couplings for the 
neutrinos but only that they be almost equal, and this may result 
from a discrete (permutation) symmetry
and its breaking.

In addition, data on neutrinos have provided clear evidence pointing towards
neutrino oscillations with very large lepton mixing between the 3 known
neutrino families \cite{data}. However, unlike in the quark sector, where
simple $S_3$ permutation symmetries can generate the general features of
quark masses and mixings, it was found \cite{jucasym} that it is impossible
to obtain large leptonic mixing angles with any general symmetry (or its
breaking) of only the 3 known generations, without having to consider the
see-saw mechanism or a more elaborate extension of the SM. Therefore, if
such symmetries exist, they must be realized in more extended scenarios.

In this Letter, we will show that it is exactly within this simple extension
of the SM (unless, as explained, one considers the see-saw mechanism or
other more elaborate extensions of the SM), just by adding a fourth
generation of SM lepton doublets and singlets, that one may obtain large
mixing angles for the leptons after breaking the symmetry.

\section{Large leptonic mixing and symmetry}

First, we consider the large leptonic mixing consistent with the present
data on neutrinos. One of the most attractive scenarios for the mixing is an
orthogonal matrix of type 
\begin{equation}
\label{f3t}F^T=\left[ 
\begin{array}{ccc}
\frac{1}{\sqrt{2}} & \frac{-1}{\sqrt{2}} & 0 \\ 
\frac 1{\sqrt{6}} & \frac 1{\sqrt{6}} & \frac{-2}{\sqrt{6}} \\ \frac 1{\sqrt{%
3}} & \frac 1{\sqrt{3}} & \frac 1{\sqrt{3}} 
\end{array}
\right] 
\end{equation}
from which we find large mixing angles, $\sin^2(2\theta _{{\rm atm}})=8/9$
for the atmospheric neutrinos, and $\sin ^2(2\theta _{{\rm sol}})=1$ for the
solar neutrinos. One may think of different frameworks, where a leptonic
mixing matrix approximate to $F^T$ arises from a simple and obvious
structure. Of course, the simplest scenario, which reproduces $F^T$, can be
obtained if one takes different structures for the charged lepton and
neutrino mass matrices. In the limit where the charged lepton mass matrix is
proportional to the democratic mass matrix $\Delta $, where all entries are
equal to $1$, while the neutrino mass matrix is proportional to ${1\>\!\!\!%
{\rm I}}$, the lepton mixing matrix will be the transpose of the orthogonal
matrix which diagonalizes $\Delta $, and this is just $F^T$ \cite{frit-yana}.

Another approach was considered in Ref. \cite{jucausy}. In the framework of
the universal strength for Yukawa couplings (USY), it was proven that large
lepton mixing is compatible with all lepton mass matrices having the same
structure, of the form 
\begin{equation}
\label{musy1}M=c\left[ 
\begin{array}{lll}
e^{i\alpha } & 1 & 1 \\ 
1 & e^{i\beta } & 1 \\ 
1 & 1 & e^{i\gamma } 
\end{array}
\right] 
\end{equation}
The hierarchy of the charged leptons requires $\alpha _e,\beta _e,\gamma _e$
to be small and $M_e$ is near to the democratic limit $\Delta $. For the
neutrino mass matrix, one must take $\alpha _\nu ,\beta _\nu ,\gamma _\nu $
to be large, otherwise the leptonic mixing angles would not be large either.
Indeed, if one takes $\alpha _\nu ,\beta _\nu ,\gamma _\nu $ to be near to $%
2\pi /3$, a lepton mixing matrix near to $F^T$ can be obtained. In the limit 
$\alpha _\nu =\beta _\nu =\gamma _\nu =2\pi /3$, the neutrino mass matrix
will be proportional to a unitary matrix $W$, and the neutrinos will be
degenerate. The mixing $F^T$ will result from a small perturbation of $W$,
just by taking $\alpha _\nu =\beta _\nu $ and $\gamma _\nu $ slightly
different from $2\pi /3$.

However, as pointed out in Ref. \cite{jucasym}, both structures, i.e. $%
M_e=\Delta $ while $M_\nu ={1\>\!\!\!{\rm I}}$ or the pattern of Eq. (\ref
{musy1}), could never be the result of a simple symmetry or of its breaking.
More precisely, there exists no symmetry, transforming the lepton fields

\begin{equation}
\label{trafo}
\begin{array}{l}
L_i\quad \rightarrow \quad P_{ij}\ L_j \\ 
e_{iR}\quad \rightarrow \quad Q_{ij}\ e_{jR} 
\end{array}
\quad {\rm or}\quad 
\begin{array}{l}
L_i\quad \rightarrow \quad P_{ij}\ L_j \\ 
e_{iR}\quad \rightarrow \quad Q_{ij}\ e_{jR} \\ 
\nu _{iR}\quad \rightarrow \quad R_{ij}\ \nu _{jR} 
\end{array}
\quad 
\end{equation}
(depending on whether neutrinos are Majorana or Dirac\footnote{%
We do not consider (heavy) mass terms for the right-handed neutrinos and the
see-saw mechanism. In fact, in the latter context, the statement we make
here is not necessarily true \cite{jucaz3}.}) and imposing invariance
conditions for the lepton mass matrices 
\begin{equation}
\label{inva1}
\begin{array}{lll}
P^{\dagger }\cdot M_e\cdot Q=M_e & \ ,\quad & P^T\cdot M_\nu \cdot P=M_\nu
\quad {\rm or}\quad P^{\dagger }\cdot M_\nu \cdot R=M_\nu , 
\end{array}
\end{equation}
that would force the charged lepton mass matrix to be proportional to the
democratic limit $\Delta $, while, at the same time, requiring that the
neutrino mass matrix be given by a matrix proportional to ${1\>\!\!\!{\rm I}}
$ or by Eq. (\ref{musy1}) with $\alpha _\nu =\beta _\nu =\gamma _\nu =2\pi
/3 $. On the contrary, it was proven that, if such a symmetry exists, the
neutrino and the charged lepton mass matrices have to be related as 
\begin{equation}
\label{relat}M_\nu ^{\dagger }M_\nu \Delta =p\Delta \quad {\rm or}\quad
M_\nu M_\nu ^{\dagger }\Delta =p\Delta \ , 
\end{equation}
and this relation\footnote{In addition, it was shown that the symmetry cannot 
prevent the neutrino mass matrix from having a part which is proportional to $\Delta$.} 
implies small mixing angles, insufficient to solve the
atmospheric neutrino problem even after the symmetry is broken.

The question then arises of whether it is possible to have a framework in
which the breaking of flavour symmetries will generate large leptonic mixing
near to $F^T$, and how. Before studying the case of 4 generations, we give
two examples in which this was achieved in the context of some elaborate
framework:

-- Huge flavour symmetries \cite{yanagi1}: a huge flavour symmetry $%
O(3)_L\times O(3)_R$ can, of course, force the (Majorana) neutrino mass
matrix to be proportional to ${1\>\!\!\!{\rm I}}$ however, the (usual)
charged lepton mass matrix will be zero. With extra Higgs fields and
appropriate mass scales it is then possible to construct a large extra term
in the Lagrangian that will result in a charged lepton mass matrix
proportional to the democratic limit $\Delta $. As explained, the mixing
will be just $F^T$.

-- See-saw mechanism: another important example, which generates large
leptonic mixing near to $F^T$, was given in Ref. \cite{jucaz3}. It was
proven that, within the see-saw mechanism, a $Z_3$ symmetry can force all
lepton mass matrices to be proportional to $\Delta $, but does not require
small mixing. It is crucial to notice that, in the see-saw mechanism, there
is an additional matrix that respects Eq. (\ref{inva1}): the heavy neutrino
Majorana mass matrix. It is this extra ingredient that prevents small
mixing. The heavy Majorana neutrino mass matrix, which is proportional to $%
\Delta $, has no (or a singular) inverse. Therefore, the effective neutrino
mass matrix can only be computed if a suitable perturbation is added. It is
then the combined perturbations of the different mass matrices that will
make it possible to a have large leptonic mixing matrix near to $F^T$.

Still, might it not be possible to obtain the same result, i.e. some
framework in which the breaking of flavour symmetries will generate large
leptonic mixing near to $F^T$, in a much simpler context? We will show that
this is indeed the case. Large leptonic mixing near to $F^T$ occurs most
naturally in the framework of the SM with just one simple extension and in
connection with a discrete symmetry. Adding a fourth generation to the SM
(making sure that the masses and mixings of the heavy extra particles
respect the experimental constraints), one can construct a simple discrete
symmetry, which will require that the neutrino as well as the charged lepton
mass matrices be proportional to a $4\times 4$ democratic mass matrix $%
\Delta $, while at the same time generating large leptonic mixings through
the breaking. 
Thus, in this case, large leptonic mixing is consistent with a (discrete) 
symmetry and its breaking.
In this sence (it is also a most simple extension of the SM) we call it "natural"; we do not violate 't Hooft's naturalness 
principle \cite{hooft} as in the case with $3$ generations, where it
was impossible, outside
the see-saw mechanism or other more elaborate extensions of the SM \cite
{jucasym}, to obtain large mixing in connection with a symmetry. 
Furthermore, we will show that it suffices to consider a small $%
Z_4$ discrete symmetry instead of the large $S_{4L}\times S_{4R}$.

\section{A fourth generation and a $Z_4$ symmetry}

Consider the SM with Dirac neutrinos\footnote{%
Again, we do not allow for a (heavy) Majorana mass term for the right-handed
neutrinos. These can be avoided, e.g. by an extra discrete symmetry where $%
\nu _{jR}\rightarrow i\nu _{jR}$, $j=1,2,3,4$. In a more complete scenario,
one must also consider an extra generation of quark doublets and up and down
singlets, e.g. in order to have anomaly cancelation.}, and one extra
generation of lepton doublets and lepton singlets: 
\begin{equation}
\label{4lag}-{L}=\lambda _{ij}^e\ \overline{L}_i\,\ \phi \,\,\
e_{jR}+\lambda _{ij}^\nu \ \overline{L}_i\,\ \tilde \phi \,\,\ \nu _{jR}+%
{\rm h.c.}\,\quad ;\quad i,j=1,2,3,4 
\end{equation}
In order to obtain mass matrices for the charged leptons and neutrinos
proportional to a $4\times 4$ democratic $\Delta $, we impose a symmetry on
the lepton fields, which is realized in following way: 
\begin{equation}
\label{trafo1}
\begin{array}{l}
L_i\quad \rightarrow \quad P_{ij}^{\dagger }\ L_j \\ 
e_{iR}\quad \rightarrow \quad P_{ij}\ e_{jR} \\ 
\nu _{iR}\quad \rightarrow \quad P_{ij}\ \nu _{jR} 
\end{array}
\quad ;\quad P={1\>\!\!\!{\rm I}}-\frac{1+i}4\ \Delta 
\end{equation}
It is easy to check that this is indeed a $Z_4$ discrete symmetry and that
the charged lepton and neutrino mass matrices must then be proportional to $%
\Delta $. Using $\Delta ^2=4\Delta $, one can verify that $P$ is unitary, $%
P^2={1\>\!\!\!{\rm I}}-\Delta /2$, $P^3=P^{\dagger }$ and $P^4={1\>\!\!\!%
{\rm I}}$. Both the neutrino and charged lepton mass matrix respect the
relation 
\begin{equation}
\label{inva2}P\cdot M\cdot P=M 
\end{equation}
From this equation it follows that 
\begin{equation}
\label{ps}M=\frac 13\left( P\cdot M\cdot P+P^2\cdot M\cdot P^2+P^3\cdot
M\cdot P^3\right) 
\end{equation}
Inserting the expressions for $P$, $P^2$and $P^3$, one finds that any mass
matrix that obeys Eq. (\ref{inva2}), and subsequently Eq. (\ref{ps}), must
fulfil the constraint 
\begin{equation}
\label{cons}M=\frac 14\left( \Delta \cdot M\cdot \Delta \right) 
\end{equation}
Finally, using the property $\Delta M\Delta =m\ \Delta $, where $m=\sum
M_{ij}$ (valid for all matrices) one obtains that $M$ has to be proportional
to $\Delta $.

At this stage, it should be mentioned that the (usual) SM charged leptons
and neutrinos of the first 3 generations all have zero mass and that the
lepton mixing for the $3\times 3$ sector is (as yet) completely arbitrary.
The unitary matrix which diagonalizes the $4\times 4$ democratic limit $%
\Delta $ is not uniquely defined and can be written as $F_4\cdot U$, where $%
U $ is an arbitrary unitary matrix that has significant elements only in the 
$3\times 3$ sector (i.e. $U_{i4}=0$, $i=1,2,3$) and 
\begin{equation}
\label{f4}F_4=\frac 12\left[ 
\begin{array}{rrrr}
-1 & 1 & 1 & 1 \\ 
1 & -1 & 1 & 1 \\ 
1 & 1 & -1 & 1 \\ 
-1 & -1 & -1 & 1 
\end{array}
\right] 
\end{equation}
Thus, it is the breaking of the $Z_4$ discrete symmetry that will give
masses to the 3 first generations and determine the lepton mixing.

By adding a small term to the democratic limit $\Delta $, we break the $Z_4$
symmetry. 
Obviously, it is possible to have different breaking patterns and 
leptonic mixing. As an example, just to illustrate the possibility of having 
large leptonic mixing near to $F^T$ we choose for the charged lepton 
and neutrino mass matrices the
following simple pattern: 
\begin{equation}
\label{pattern}M_{e,\nu }=k_{e,\nu }\ \left( \Delta +P_{e,\nu }\right) 
\end{equation}
where%
$$
P_e={\rm diag}(0,a_e,b_e,c_e)\quad ;\quad P_\nu =\left[ 
\begin{array}{llll}
0 & 0 & 0 & 0 \\ 
0 & a_\nu & 0 & a_\nu \\ 
0 & 0 & b_\nu & b_\nu \\ 
0 & a_\nu & b_\nu & b_\nu +a_\nu 
\end{array}
\right] \, 
$$
For simplicity, we take real parameters. There is not yet any (or no) 
experimental evidence for  
CP violation in the lepton sector.
The hierarchy of the charged leptons requires that $a_e\ll b_e\ll c_e\ll 1$.
From the neutrino hierarchy $\Delta m_{21}^2\ll \Delta m_{32}^2$ we obtain
also $a_\nu \ll b_\nu \ll 1$. In addition, we get the relations $m_{\nu
_2}^2\approx \Delta m_{21}^2$and $m_{\nu _3}^2\approx \Delta m_{32}^2$
because of the restricted number of parameters for the neutrino mass matrix
(and $a_\nu \ll b_\nu $). From the invariants, ${\rm tr}(M)$, $\chi _2(M)$, $%
\chi _3(M)$ and $\det (M)$ for the charged lepton and neutrino mass matrix,
it is easy to derive the following first order approximations: 
\begin{equation}
\label{experi}
\begin{array}{l}
\begin{array}{clll}
k_e=\frac 14m_{\ell _4}, & a_e=\frac{8m_e}{m_{\ell _4}}, & 
|b_e|=\frac{6m_\mu }{%
m_{\ell _4}}, & c_e= 
\frac{16m_\tau }{3m_{\ell _4}} \\  &  &  &  
\end{array}
\\ 
\begin{array}{clll}
k_\nu =\frac 14m_{_4}, & a_\nu =\frac{4m_2}{m_{_4}}, & b_\nu =\frac{4m_3}{m_4}%
, &  
\end{array}
\end{array}
\end{equation}
where $m_{\ell _4}$ and $m_{_4}$ are the masses of the (extra) fourth
charged lepton and neutrino. At tree level, one neutrino is massless, $m_1=0$%
. In a first order approximation, we find $(a_\nu /b_\nu )^2=\Delta
m_{21}^2/\Delta m_{32}^2$ and $m_{\nu _2}=(\Delta m_{21}^2)^{1/2}$, $m_{\nu
_3}=(\Delta m_{32}^2)^{1/2}$.

The hierarchy of the parameters of the neutrino and charged lepton mass
matrices allows for a straightforward first order computation of the
orthogonal matrices that diagonalize $M_e$ and $M_\nu $. After a weak basis
transformation to a heavy basis, where the $\Delta $ part in Eq. (\ref
{pattern}) becomes diagonal, i.e. $M_{e,\nu }\rightarrow F_4^T\cdot M_{e,\nu
}\cdot F_4$, we need only find orthogonal matrices that will eliminate, from
the first two horizontal lines of $M_e$ and $M_\nu $, the parameter that
will be next in the order. So, from the first horizontal line of $M_e$ we
eliminate $b_e$ and $c_e$, and from the second line we eliminate $c_e$. For
the neutrino mass matrix, as the first eigenvalue is zero, we eliminate only 
$b_\nu $ from the second line. Thus, we obtain in approximation for the
matrices that diagonalize $M_e$ and $M_\nu $ of our ansatz in Eq. (\ref
{pattern}): 
\begin{equation}
\label{unit}U_e=F_4\cdot F_3\quad ,\quad U_\nu =F_4 
\end{equation}
where $F_3$ is the $4\times 4$ orthogonal matrix that contains, in the $%
3\times 3$ SM sector, exactly our original $F$ (its transpose was given in
Eq. (\ref{f3t})), and $(F_3)_{i4}=0$, $i=1,2,3$. As a final step, to obtain
the total orthogonal matrices that diagonalize $M_e$ and $M_\nu $ we have to
multiply $U_e$ and $U_\nu $ by matrices $I_e$ and $I_\nu $ which are very
near to ${1\>\!\!\!{\rm I}}$. For the charged leptons $(I_e)_{ij}=O(m_{\ell
_i}/m_{\ell _j})$ for $i<j$, and for the neutrinos the angles are almost
insignificant. Therefore, in a first order approximation, the lepton mixing
matrix will be 
$$
V=U_e^T\cdot U_\nu =F_3^T\ , 
$$
which coincides exactly with our original $F^T$ for the first 3 generations
and the $3\times 3$ SM sector. The mixing with the fourth generation is at
most $O(m_\tau /m_{\ell _4})$ between the 3rd and 4th families. We give a
numerical example.

Input: 
\begin{equation}
\label{input}
\begin{array}{l}
\begin{array}{cll}
a_e=4.11\times 10^{-5} & b_e=-6.34\times 10^{-3} & c_e=9.66\times 10^{-2}
\end{array}
\\ 
\begin{array}{cl}
a_\nu =6.0\times 10^{-13} & b_\nu =5.1\times 10^{-12}
\end{array}
\end{array}
\end{equation}
For the heavy extra charged lepton and neutrino, we have chosen masses%
\footnote{%
These masses are also in accord with the values given by Novikov, Okun,
Rozanov and Vysotsky\cite{novi}} $m_{\ell _4}\ =100\ {\rm GeV}$ and $m_4\
=50\ {\rm GeV}$.

Output: 
\begin{equation}
\label{output}
\begin{array}[b]{l}
\begin{array}{ccc}
m_e\ =\ 0.511\ {\rm MeV}\ ,\quad & m_\mu \ =\ 105.7\ {\rm MeV},\quad & 
m_\tau \ =\ 1777\ {\rm MeV} 
\end{array}
\\ 
\begin{array}{cc}
\Delta m_{21}^2\ =\ 5.6\times 10^{-5}\ {\rm eV}^2 ,\quad & \Delta m_{32}^2\
=\ 4.0\times 10^{-3}\ {\rm eV}^2 , 
\end{array}
\\ 
m_{\nu _3}\ =0.064\ {\rm eV} 
\end{array}
\end{equation}
and

\begin{equation}
\label{eq22}|V|=\ \left[ 
\begin{array}{cccc}
0.706 & 0.708 & 2.14\times 10^{-3} & 7.3\times 10^{-6} \\ 
0.422 & 0.418 & 0.805 & 8.6\times 10^{-4} \\ 
0.569 & 0.569 & 0.594 & 1.1\times 10^{-2} \\ 
6.5\times 10^{-3} & 6.5\times 10^{-3} & 5.7\times 10^{-3} & 0.9999
\end{array}
\right] \ 
\end{equation}
from which we obtain 
\begin{equation}
\label{eq23}
\begin{array}{cc}
\sin ^2(2\theta _{{\rm atm}})\ =\ 0.913,\quad  & \sin ^2(2\theta _{{\rm sol}%
})\ =\ 1.0
\end{array}
\end{equation}
The value for $|V_{\tau 4}|^2=1.17\times 10^{-4}$ is above the value permitted 
by the data 
available from DELPHI \cite{data1} where $|V_{\tau 4}|^2<3\times 10^{-5}$ for 
$m_{\nu_4}=50 \rm{GeV}$. The values for the atmospheric and solar neutrino 
mixing are in good agreement with the values for the Large Mixing Angle (LMA) 
scenario from previous analysis. At present, 
slightly different values, where 
$\sin ^2(2\theta _{{\rm atm}})$ is almost maximal, near to $1.0$, 
and $\sin ^2(2\theta _{{\rm sol}}$ is 
somewhat smaller than $1$, seem to be favoured. However, it is not yet 
completely clear what the parameter range is for the neutrino masses and mixings.
In any case, both scenarios can be accommodated in the scheme presented here.
\section{Conclusions}

We have shown that large leptonic mixing occurs most naturally in the
framework of the SM, just by adding a fourth generation to the SM. One can
then construct a very simple discrete symmetry that will require the
neutrino as well as the charged lepton mass matrices be proportional to a $%
4\times 4$ democratic mass matrix $\Delta $. Without considering the see-saw
mechanism, or other more elaborate extensions of the SM, and contrary to the
case with only 3 generations, large leptonic mixing is obtained when the
symmetry is broken. Furthermore, it was shown that it suffices to consider a
small $Z_4$ discrete symmetry instead of the large $S_{4L}\times S_{4R}$.

\subsection*{Acknowledgements}I am grateful to G.C. Branco for suggestions. This work received partial
support from the Portuguese Ministry of Science - Funda\c c\~ao para a
Ci\^encia e Tecnologia.

\end{document}